\documentclass[conference]{IEEEtran}

\pdfoutput=1
\usepackage{etex}

\usepackage[utf8]{inputenc}
\usepackage[english]{babel}
\usepackage[bookmarks=false,hidelinks]{hyperref}
\usepackage{balance}
\usepackage{tabularx}
\usepackage{xcolor}
\usepackage[autolanguage]{numprint}
\usepackage{xspace}
\usepackage{graphicx}
\usepackage{multirow}
\usepackage{booktabs}
\usepackage[inline]{enumitem}

\AtBeginDocument{%
}

\newcommand{\answer}[2]{\vspace{.3cm}{\centering\fbox{\parbox{0.97\columnwidth}{\textbf{RQ#1}. #2}}}\vspace{.1cm}}

\newcommand{\travis}[0]{Travis~CI\xspace}

\newcommand{\nbConfigChange}[0]{\numprint{709220}\xspace}
\newcommand{\nbConfigChangePercent}[0]{\numprint{7.34}\%\xspace}
\newcommand{\nbProjoectChangeConfig}[0]{\numprint{104708}\xspace}
\newcommand{\nbProjoectChangeConfigPercent}[0]{\numprint{38.36}\%\xspace}

\newcommand{\nbProjects}[0]{\numprint{272917}\xspace}
\newcommand{\nbUsers}[0]{\numprint{123168}\xspace}
\newcommand{\nbJobs}[0]{\numprint{35793144}\xspace}
\newcommand{\startTravisData}[0]{30 September 2018\xspace}
\newcommand{\travisDataEnd}[0]{22 January 2019\xspace}

\usepackage{amssymb,amsmath}
\usepackage{ifthen}
\newboolean{showcomments}
\setboolean{showcomments}{true}
\ifthenelse{\boolean{showcomments}}
 { \newcommand{\mynote}[2]{
      \fbox{\bfseries\sffamily\scriptsize#1}
        {\small$\blacktriangleright$\textsf{\emph{#2}}$\blacktriangleleft$}}}
        { \newcommand{\mynote}[2]{}}

\begin{document}

\title{An Analysis of 35+ Million Jobs of Travis CI}

\author{
    \IEEEauthorblockN{Thomas Durieux\IEEEauthorrefmark{1},
Rui Abreu\IEEEauthorrefmark{1},
Martin Monperrus\IEEEauthorrefmark{3},\\
Tegawendé F. Bissyandé\IEEEauthorrefmark{2},
Luís Cruz\IEEEauthorrefmark{4}},

\IEEEauthorblockA{
 \IEEEauthorrefmark{1}INESC-ID and IST, University of Lisbon,
 \IEEEauthorrefmark{3}KTH Royal Institute of Technology,\\
 \IEEEauthorrefmark{2}University of Luxembourg,
 \IEEEauthorrefmark{4}INESC-ID and University of Porto
}
}

\maketitle

\begin{abstract}
    \travis handles automatically thousands of builds every day to, amongst other
    things, provide valuable feedback to thousands of open-source developers.
    In this paper, we investigate \travis to firstly understand who is using it, and when they start to use it.
    Secondly, we investigate how the developers use \travis and finally, how frequently the developers change the \travis configurations.
    We observed during our analysis that the main users of \travis are corporate users such as Microsoft. 
    And the programming languages used in \travis by those users do not follow the same popularity trend than on GitHub, for example, Python is the most popular language on \travis, but it is only the third one on GitHub. 
    We also observe that \travis is set up on average seven days after the creation of the repository and the jobs are still mainly used (60\%) to run tests.
    And finally, we observe that 7.34\% of the commits modify the \travis configuration.
    We share the biggest benchmark of \travis jobs (to our knowledge): it contains \nbJobs jobs from \nbProjects different GitHub projects.
\end{abstract}

\section{Introduction}

In the last years, more and more manual software engineering tasks have been assisted or replaced by automatization processes.
One of the most popular automatizations in software engineering is continuous integration.
This concept of continuous integration was initially to ensure that the new commits are correctly integrated inside the software, i.e., for a new commit or at a specific scheduled time interval, the tests are executed to identify regression bugs in the applications \cite{fowler2006continuous}.
However, this concept evolved with time, and it is no longer limited to building and testing applications.
Indeed, continuous integration is now used for new usages such as code analysis and application deployments. This evolution is illustrated with the new features are proposed the continuous integration tools, such as automatic deployment.

However, those new usages are little studies.
It is crucial to analyze those usages since continuous integration is more and more used and taught.
Additional knowledge is mandatory to understand the requirements, difficulties of the developers, and to be able to provide new solutions to improve their workflow and new needs.

In this paper, we contribute to this vision by studying the integration of the continuous integration in open-source repositories.
We investigate the biggest continuous integration success story \cite{Beller2018BlockchainbasedSE}: \travis.
\travis is the most popular open-source continuous integration service for GitHub.
We consider different aspects of understanding the usage of \travis.
Firstly, we study who is using \travis, secondly when developers integrate \travis in their projects, then we analyze the different usages that developers have on \travis and finally, we look for if the developers are maintaining their automatization environments.

To sum up, our contributions are:
\begin{itemize}
\item An analysis of \travis that targets four aspects: who use it, when the developers start to use it, for which purposes they use it and how \travis configuration evolves. Those aspects are novel compared to the closest related work \cite{beller2017travistorrent,Beller:2017:OMT:3104188.3104232}.
\item A benchmark of all \travis jobs executed during \startTravisData to \travisDataEnd. It contains \nbJobs \travis jobs triggered by \nbProjects projects. The benchmark is available on Zenodo with the DOI: \href{https://doi.org/10.5281/zenodo.2560966}{10.5281/zenodo.2560966} 
for future research. The tool-set that has been used to create the benchmark is available on GitHub.\footnote{The tool-set to collect to create the benchmark: 
\url{https://github.com/tdurieux/travis-listener}
} For comparison, Hilton et al. \cite{hilton2016usage}'s study considers \numprint{12000} projects, our dataset has data from \numprint{250000} projects.
\end{itemize}


\autoref{sec:background} presents what is \travis.
\autoref{sec:analysis} presents our analysis of \travis in four research questions.
\autoref{sec:rw} presents the related works of this study and \autoref{sec:conclusion} concludes this paper.

\section{What is \travis?}\label{sec:background}

\travis is a company that offers an open-source continuous integration service that is tightly integrated with GitHub.
It allows developers to build their projects without maintaining their own infrastructure.
\travis provides a simple interface to configure build tasks that are executed for a set of given events: pull requests, commits, crons, and API calls.
Currently, \travis supports 34 different programming languages including Python, NodeJS, Java, C, C++ in three different operating systems: Linux, Windows, and Mac OSX.
It also provides additional services that support, for example, Docker, Android apps, iOS apps, and databases.
The \travis service is free for open-source projects, and a paid version is available for private projects.
It is currently used by more than \numprint{932977} open-source projects and \numprint{600000} users.\footnote{From \url{https://travis-ci.org}, visited \today{}}

\begin{figure}[t]
    \centering
    \includegraphics[width=0.9\columnwidth]{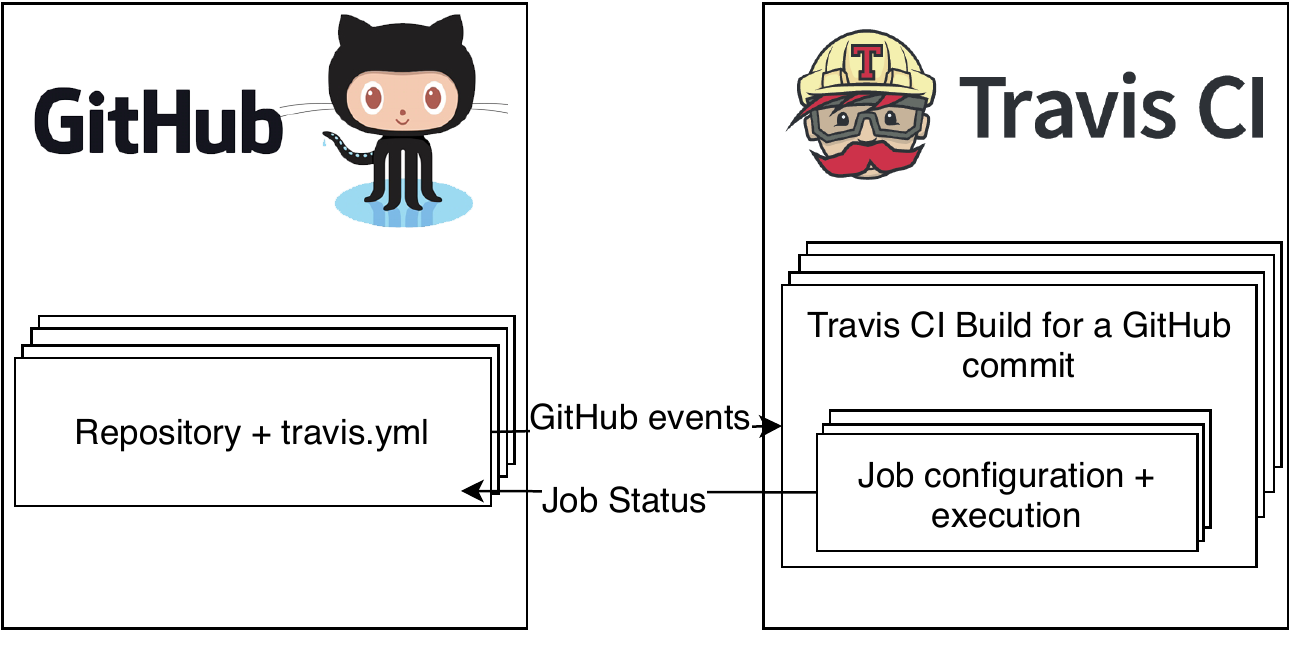}
    \caption{Architecture of \travis and its integration with GitHub.}\label{fig:travis}

\end{figure}

\autoref{fig:travis} presents a high level representation of \travis infrastructure.
\travis interacts with GitHub with a set of \underline{webhooks} that are triggered by GitHub events.
For each event, \travis sets up a new \underline{build} by reading the configuration that the developers wrote in their repository (.travis.yml file).
Each build is composed of one or several jobs.
A \underline{job} is the execution of the build in a specific environment, for example, one job runs with Java 8 and one with Java 9, or a job can also be used for specific tasks such as deploying Docker images.
According to our benchmark, on average, each build contains 3.72 jobs.

\section{\travis Analysis}\label{sec:analysis}

In this section, we present our study on \travis to understand the behavior of the developers regarding the automatization of their open-source repositories.

\subsection{Research Questions}

To achieve the goal of this analysis, we focus on four different aspects:

\begin{itemize}[leftmargin=*]
    \item \textbf{RQ1}. Who is using \travis? This first research question aims to identify which type of users or programming communities use \travis and at which scale.
    \item \textbf{RQ2}. Do the projects use \travis since their inception? In this research question, we analyze how much time the developers take to setup \travis in their projects, and we observe if there is a different behavior depending on the type of user.
    \item \textbf{RQ3}. How is \travis used? The next question is to understand to what extent \travis is used to execute tasks that are not related to testing.
    \item \textbf{RQ4}. To what extent do \travis configurations evolve over time? The final question studies the evolution of the \travis project configuration in order to understand if the developers take care of maintaining their build configurations.
\end{itemize}

\subsection{Study Design}

\begin{table}[t]
    \centering{}
    \caption{The main statistics of our benchmark}\label{tab:benchmark}
    \begin{tabularx}{0.48\textwidth}{@{}X r@{}}
        \toprule
        \# Job execution & \nbJobs \\
        \# Projects & \nbProjects \\
        \# Users & \nbUsers \\
        \# Period of the study & \startTravisData to  \travisDataEnd\\
        \bottomrule
    \end{tabularx}
\end{table}

To answer our research questions, we create a new benchmark with data extracted from \travis and GitHub.
We agnostically collected all job information of \travis from the \startTravisData to the \travisDataEnd.
The main statistics of the benchmark are presented in \autoref{tab:benchmark}.
During that period, we collected \nbJobs build jobs, which represent \numprint{59}G of raw data.
In addition to those job configurations from \travis, we collected the GitHub data related to the repositories that use \travis during the studied period.
We collected data from  \nbProjects different repositories which represent \numprint{2.3}G of raw data. The benchmark is available on Zenodo with the DOI: \href{https://doi.org/10.5281/zenodo.2560966}{10.5281/zenodo.2560966} 
for future research. The tool-set that has been used to create the benchmark is available on GitHub: 
\url{https://github.com/tdurieux/travis-listener}.

\subsection{RQ1. Who is currently using \travis?}

The first research question that we investigate is to understand the types of user that use \travis.
We first look at the number of fork repositories that use \travis compared to non-fork repositories. 
Then we look at the number of organization and individual accounts that use \travis.

\begin{table}
    \centering{}
    \caption{Proportion of fork/non-fork in \travis.}
    \label{tab:statFork}
    \begin{tabularx}{0.48\textwidth}{@{}X c c@{}}
    \toprule
                & Fork  & Non-Fork \\ \midrule
    \# Projects & \numprint{30579} (11.56\%) & \numprint{233880} (88.43\%)\\
    \bottomrule
    \end{tabularx}
\end{table}

The fork metric indicates the number of forks that are used for active development on GitHub.
Indeed, setting up \travis on a fork is an additional step that only active developers do. 
There are two use cases: the first use-case is a developer that frequently contributes to a project using pull requests and wants to ensure the correct behavior of her code before opening the pull request. 
The second use-case is that the developers that fork a repository to continue or change the direction of the project.
\autoref{tab:statFork} presents the results of this study. It shows that most of the active \travis users are working on non-forked repositories, and 11.56\% of the repositories are forks.
It indicates as expected that non-fork repositories are bigger \travis users, but sill \numprint{30579} forked repositories used \travis during the studied period. 
It means that the owner of those repositories did the additional step to increase the quality of their contributions.

\begin{table}[t]
    \centering{}
    \caption{Proportion of user vs. organization in \travis.}
    \label{tab:statUser}
    \begin{tabularx}{0.48\textwidth}{@{}X c c@{}}
    \toprule
                & Individual User    & Organization\\ \midrule
    \# Projects & \numprint{158446} (59.91\%) & \numprint{106013} (40.08\%)\\
    \bottomrule
    \end{tabularx}
\end{table}

The second metric is related to the number of users and organizations that use \travis.
This metric reflects if an organization is more likely to set up \travis compare to traditional users.
\autoref{tab:statUser} shows the number of repositories that are owned by individual users vs. organizations, according to GitHub API.
It shows that 59.91\% of the repositories are owned by individual users.
However, considering the number of repositories owned by organization vs. users (31 million users vs. 2.1 million organizations\footnote{GitHub statistics: \url{https://octoverse.github.com/} (visited 15 June 2019)}) it is much likely that an organization that owns a repository will setup \travis compared to individual users.
The organizations are as expected the biggest \travis users in term of jobs executed.
\autoref{tab:topusers} presents the top 10 biggest users of \travis.
Those ten users represent 4.06\% (\numprint{1453755} jobs) of the total amount of jobs executed by \travis.
The new owner of GitHub (Microsoft + Azure) is the biggest \travis user, followed by the Apache foundation, Elastic and Mozilla. We note the absence of the other big software companies such as Google, Apple, Facebook, or Amazon.
\begin{table}
    \centering{}
    \caption{The Biggest \travis users.}
    \label{tab:topusers}
    \begin{tabularx}{.8\columnwidth}{@{}r X r r@{}}
    \toprule
        \# & Owner & \# Jobs & \# Projects \\\midrule
        1 & Apache & \numprint{248154} & \numprint{262} \\
        2 & Elastic & \numprint{188757} & \numprint{38} \\
        3 & Mozilla & \numprint{168394} & \numprint{174} \\
        4 & Azure & \numprint{161169} & \numprint{143} \\
        5 & Microsoft & \numprint{151654} & \numprint{225} \\
        6 & Robertdebock & \numprint{129121} & \numprint{91} \\
        7 & Rust-lang & \numprint{115240} & \numprint{29} \\
        8 & Rails & \numprint{111530} & \numprint{26} \\
        9 & Mike-north & \numprint{94423} & \numprint{52} \\
        10 & Pytorch & \numprint{85313} & \numprint{8} \\
    \bottomrule
    \end{tabularx}
\end{table}

The final observation about who is using \travis is about the language that the developers use in \travis compared to the languages that they use in GitHub.
\autoref{tab:travislanguage} presents the most popular languages of \travis and compares them to GitHub ranking\footnote{GitHub language ranking: \url{https://github.blog/2018-11-15-state-of-the-octoverse-top-programming-languages/}}, \textit{N.A.} is used when the language is not present in the top 10 of GitHub.
We observe that the \travis popular language is uncorrelated with the ranking of GitHub. 
This shows that some language communities, such as Python, PHP, C, Go, Rust, have a stronger usage of \travis compared to their popularity.
It seems to indicate that those language environments have a stronger culture of continuous integration compared to other environments.

\begin{table}[t]
    \centering{}
    \caption{Most popular language on \travis compared to GitHub. \textit{N.A.} is used when the language is not present in the top 10 of GitHub}
    \label{tab:travislanguage}
    \begin{tabularx}{.8\columnwidth}{@{}r r X r@{}}
        \toprule
        \multicolumn{2}{c}{Rank} & & \\
        \travis & GitHub & \multirow{-2}*{Programing language} & \multirow{-2}*{\# Builds}\\\midrule
        1 & 3 & Python & \numprint{7793364} \\
        2 & 1 & NodeJs & \numprint{6441830} \\
        3 & 4 & PHP & \numprint{3387538} \\
        4 & 10 & Ruby & \numprint{3030574} \\
        5 & 5 & C++ & \numprint{2799603} \\
        6 & 9 & C & \numprint{2459281} \\
        7 & 2 & Java & \numprint{2200925} \\
        8 & \textit{N.A} & Go & \numprint{1512233} \\
        9 & 8 & Shell & \numprint{1461724} \\
        10 & \textit{N.A} & Rust & \numprint{1054800} \\
        \bottomrule
    \end{tabularx}
\end{table}


\answer{1}{
\textbf{Who is using \travis?}
Our experiment reveals that \travis is used by a large diversity of users, by more than  \nbUsers unique users uses \travis during the studied period.
Moreover, the biggest \travis users are corporate institutions that have open-source projects such as Elastic Search or Microsoft.
This study also reveals that some programming language communities are more active on \travis than others, for example, Python is the most popular language on \travis but is only the third over GitHub repositories.}

\subsection{RQ2. Do the projects use \travis since their inception?}

In the second research question, we investigate when \travis users start to use \travis in order to understand the habit of the developers.
In this study, we consider that a project uses \travis since the beginning when the first \travis job is started in the 48 hours after the creation of the GitHub repository.
We investigate firstly if the type of project and user has an impact on the \travis setup time.
Secondly, we analyze how the setup time evolves with the age of the project.

\begin{table}[t]
    \centering{}
    \caption{Projects that start to use \travis within 48 hours after the creation of the GitHub repository.}
    \label{tab:timeProject}
    \begin{tabular}{@{}l|l|r|r@{}}
    \toprule
        Project type & User type & \# Projects & \% \\ \midrule
        Fork & Individual user & \numprint{10369} & 40.05\% \\
        Fork & Organization & \numprint{1504} & 32.06\% \\
        Non-fork & Individual user & \numprint{59411} & 44.81\% \\
        Non-fork & Organization & \numprint{33886} & 33.44\% \\
    \bottomrule
    \end{tabular}
\end{table}

To understand the topology of the setup of \travis, we first look if the type of project and the type of user have an impact on the setup time of \travis.
\autoref{tab:timeProject} presents the number of projects that start to use \travis in the 48 hours of their creation.
We can observe that individual users set up more frequently \travis since the beginning compared to organization projects.
However, there is no major difference between forked projects and non-forked projects.
A potential explanation of the difference between individual users and organizations can be that projects from organizations are started internally before being released publicly. Consequently, the first \travis build will be when the project is made available instead of when the first commit is pushed.

\begin{table}[t]
    \centering{}
    \caption{Average and median time use by the developers to setup \travis depending on the year of creation.}
    \label{tab:timeAge}
    \begin{tabular}{@{}l l l r r@{}}
    \toprule
    Creation year & Average & Median  & \# Projects & \% \\ \midrule
    2008 & 5.2 years  & 4.85 years & \numprint{213} & 0.07\% \\
    2009 & 4.93 years  & 4.59 years & \numprint{621} & 0.22\% \\
    2010 & 3.91 years  & 3.64 years & \numprint{1461} & 0.53\% \\
    2011 & 3.04 years  & 2.75 years & \numprint{3295} & 1.2\% \\\midrule
    \multicolumn{5}{c}{\travis creation} \\\midrule
    2012 & 2.29 years  & 1.99 years & \numprint{5625} & 2.06\% \\
    2013 & 1.6 years  & 1.08 years & \numprint{9945} & 3.64\% \\
    2014 & 1.08 years  & 6.56 months & \numprint{16139} & 5.91\% \\
    2015 & 8.77 months  & 2.38 months & \numprint{25349} & 9.28\% \\
    2016 & 5.46 months  & 24.85 days & \numprint{36728} & 13.45\% \\
    2017 & 2.9 months  & 8.18 days & \numprint{54170} & 19.84\% \\
    2018 & 23.3 days  & 1.44 days & \numprint{102644} & 37.6\% \\
    2019 & 1.37 days  & 2.18 hours & \numprint{8269} & 3.02\% \\ \midrule
    Total   & 5.55 months & 7.6 days  & \nbProjects & 100\%\\
    \bottomrule
\end{tabular}
\end{table}

\autoref{tab:timeAge} presents the results of our second investigation regarding the setup time.
The table contains the average and median time of the setup of \travis depending on the age of the project.
The first column presents the age of the project; the second and third columns present the average and median time for setting up \travis.
Finally, the two last columns present the number of projects created for a given year and the proportion of the total number of studied projects.

We observe that the setup time of \travis is decreasing with time.
In the year of \travis creation, it takes almost two years for the projects to set up \travis, and nowadays in 2019, the median time is 2.18 hours.
This change can be explained firstly by the increasing popularity of \travis,
secondly by the \travis GitHub app that automatically sets up \travis when a repository is created.

\answer{2}{
    \textbf{Do the projects use \travis since their inception?}
    We observed that it takes seven days (median) to set up \travis in a GitHub repository. We also noticed that older projects take months up to years, but this time has drastically decreased over the last two years.
    Individual users are more prone to set up \travis compare to organizations.
}

\subsection{RQ3. How is \travis used?}

Now that we have a better understanding of who and when \travis is used by developers, we study the usage of \travis by the developers.
The goal is to identify the different usages and to what extent they are used.

In order to achieve this goal, we manually analyzed build configurations and commit messages to identify categories of usage.
Then, we select keywords that we use to classify automatically the \nbJobs \travis jobs that we collected.

We identify the following eight categories:
\begin{enumerate}
    \item {[Building]} building jobs are used to compile and verify that the project still compiles;
    \item {[Testing]} testing jobs are used to compile and run the test suite of the application;
    \item {[Releasing]} releasing jobs are used to deploy the project binaries or the docker images;
    \item {[Analyzing]} analyzing jobs are performing static analysis to detect bugs, typos or to assess the code quality of the project;
    \item {[Formatting]} formatting jobs verify that the source code is correctly formatted or that the license headers are correctly placed;
    \item {[Documentation]} documentation jobs are used to deploy documentation or websites of the project;
    \item {[Communication]} communication jobs consist of communicating information to the developers using email, Slack or GitHub comments and
    \item {[Unknown]} the final category contains the job that we did not succeed to categorize.
\end{enumerate}

\autoref{tab:usageCategory} presents the results of the classification.
The first column contains the usage category, the second column contains the number of jobs present in that category, and the final column presents the proportion of this category over the complete benchmark.

The main observation is that the testing and building are the most frequent usage in \travis with 66.94\% of the usage.
Those two usages are followed by the documentation and formatting usages with 3.96\% and 1.82\% of the jobs, respectively.
Those results show that the developers use \travis for other purposes than traditional testing; however, this usage is still marginal compared to testing.
We plan to reproduce this experiment in one year to observe the evolution of the usage in \travis.

\begin{table}[t]
    \centering{}
    \caption{The number of jobs for each usage category.}
    \label{tab:usageCategory}
    \begin{tabularx}{0.48\textwidth}{@{}X r r@{}}
        \toprule
        Usage & \# Jobs & \% \\ \midrule
        Testing & \numprint{20991572} & 58.64\% \\
        Building & \numprint{2973544} & 8.30\% \\
        Documentation & \numprint{1170264} & 3.26\% \\
        Formatting & \numprint{653291} & 1.82\% \\
        Releasing & \numprint{514107} & 1.43\% \\
        Analyzing & \numprint{64896} & 0.18\% \\
        Communication & \numprint{26059} & 0.07\% \\\midrule
        Unknown & \numprint{9399411} & 26.26\% \\
        \bottomrule
    \end{tabularx}
\end{table}

\answer{3}{
    \textbf{How is \travis used?}
    According to our analysis, \travis is still mainly used for traditional building and testing activities.
    However, more than two millions of jobs are dedicated to other usages such as documentation deployment, code analysis, and code formatting.
    It shows that developers are now considering continuous integration for other purposes.
}

\subsection{RQ4. To what extent, \travis configurations evolve over time?}

The previous research question focuses on the different usages of \travis.
Now, in this research question, we analyze how frequently developers change their \travis configurations.
This frequency shows the interest of the developers to maintain their configuration in a working state or the difficulty to set up the \travis environment.

The methodology that we follow to track those changes, is the look at the \travis configuration of each job and follow any change in their configuration.
We track the configuration for each project but separating the configuration for each repository branch and for each build environment (called build matrix in \travis). Once we detect a change, we collect the commit SHA that triggered the job and finally count the unique commits that change the \travis configurations.

Following this methodology, we observe that \nbConfigChange commits (\nbConfigChangePercent) change the configuration during the studied period.
Only \nbProjoectChangeConfig projects (\nbProjoectChangeConfigPercent) change their configuration.
The results indicate that the majority of the projects have a stable configuration.

We manually analyze a sample of commits that change the configuration, and we observe that a significant number of builds are related to debugging the \travis configuration.
It appears that the developers have trouble to set up a stable environment, especially when they are dealing with complex environments such as building mobile applications.

\answer{4}{
    \textbf{To what extent do \travis configurations evolve over time?}
    We observe that \nbConfigChangePercent of commits modify \travis configurations. This is the first experimental report of developers modifications of CI configuration.
    Further empirical studies are needed to understand the evolution of CI configuration better.
}

\section{Related Works}\label{sec:rw}

Beller et al. \cite{beller2017travistorrent} present TravisTorrent a benchmark of \travis builds where information is extracted from \travis and GitHub such as the number of builds, the message of the associated PR.
The difference between TravisTorrent and our benchmark is that the age of the data and the completeness of the collected data. 
Indeed, TravisTorrent focuses on specific repositories (\numprint{1300}), in our benchmark we collected all the \travis jobs between \startTravisData and \travisDataEnd.
Their following study \cite{Beller:2017:OMT:3104188.3104232} exploits this benchmark to study the build behavior of the projects that use \travis.
Compared to this paper, we focus our analysis on different aspects. They focused on the outcome of the builds, and we focus on the usage and evolution of the \travis configuration.

Hilton et al. \cite{hilton2016usage} study the use of continuous integration in open-source projects.
It shows that continuous integration has a positive impact on the projects, and it is used in 70\% of the most popular projects on GitHub.
In this paper, we study a different aspect of continuous integration as well as including a larger number of builds and projects.

Zhao et al. \cite{zhao2017impact} study the impact of \travis on development practices. Their main finding is that GitHub pull requests are more frequently closed after the integration of \travis.
We did not focus our investigation on the impact of \travis on the development practices, but we focus on who, how, and for which purpose \travis is used.


Rausch et al. \cite{rausch2017empirical} present a study on 14 open-source projects that use GitHub and \travis. 
They analyzed the build failure and identified 14 different error categories.
They presented several seven observations such as: ``authors  that  commit less frequently tend to cause fewer build failures'', or ``Build failures mostly occur consecutively''.
We focused our analysis on the usage of \travis and did not analyze the outcome of the build.
Moreover, we consider a much higher number of projects compared to this work.

Widder et al. \cite{widder2018m} present a study that analyzes the reasons why projects are leaving \travis. 
They observed that this phenomenon is related to the build duration and the repository language.
They showed that C\# repositories are more likely to quit \travis because \travis did not support Windows virtual machine (\travis nowadays supports Windows virtual machines). 
On the contrary, repositories that have long build are more likely to continue to use \travis.
In this paper, we focused our analysis on the usage and did not consider the evolution of the usage over time since we only focus on a four months period.
It would be interesting to reproduce the experiment of this paper on the data from 2018 to 2019 when \travis supports Windows virtual machines.

\section{Conclusion}\label{sec:conclusion}

In this study, we analyzed the developer's usages of \travis, one of the most popular build system.
We collected \nbJobs \travis jobs from \nbProjects projects and
we observe that \travis is more and more popular and developers on GitHub uses it more rapidly.
It is as much used by big companies than individual users (40\% vs. 60\%) that care about the status of their builds.
Indeed, \nbConfigChangePercent of the commits that trigger \travis changes the build configuration.
Testing and building project are still the most popular usages, but new usages such as deploying documentation and websites, code analysis, and formatting start to emerge on \travis.
And in 2019, developers take only on average 1.37 days to set up \travis.
It shows that the developers are interested in the automatization systems, and they are using CI for other tasks than pure project testing and building.

\section*{Acknowledgments}

This work was supported by Fundação para a Ciência e a Tecnologia (FCT), with the reference PTDC/CCI-COM/29300/2017.

\bibliographystyle{IEEEtran}
\bibliography{references}

\balance

\end{document}